\documentstyle[prl,floats,aps,twocolumn,epsf,graphicx]{revtex}
\begin{document}
\twocolumn[\hsize\textwidth\columnwidth\hsize\csname
@twocolumnfalse\endcsname

\title{Strategy updating rules and strategy distributions in dynamical multiagent systems}
\author{Shahar Hod$^1$ and Ehud Nakar$^2$}
\address{$^1$Department of Condensed Matter Physics, Weizmann Institute, Rehovot 76100, Israel}
\address{}
\address{$^2$The Racah Institute of Physics, The
Hebrew University, Jerusalem 91904, Israel}
\date{\today}
\maketitle

\begin{abstract}

\ \ \ In the evolutionary version of the minority game, agents update 
their strategies (gene-value $p$) in order to improve their performance. 
Motivated by recent intriguing results obtained for prize-to-fine ratios which are smaller than unity, 
we explore the system's dynamics with a strategy updating rule of 
the form $p \to  p \pm \delta p$ ($0 \leq p \leq 1$). We find that the strategy distribution 
depends strongly on the values of the prize-to-fine ratio $R$, the length scale $\delta p$, and the type of 
boundary condition used. We show that these parameters determine the amplitude and frequency of the 
the temporal oscillations observed in the gene space. 
These regular oscillations are shown to be the main factor which determines the strategy distribution of the population. 
In addition, we find that 
agents characterized by $p={1 \over 2}$ (a coin-tossing strategy) have the best chances of 
survival at asymptotically long times, regardless of the value of $\delta p$ and the boundary conditions used.
\end{abstract}
\bigskip

]

The Minority Game (MG) is a successful model describing a population of competing and evolving individuals. 
This complex system has been explored extensively in the last few years, 
see e.g., \cite{HubLuk,ChaZha,DhRo,SaMaRi,John1,Cev,BurCev,LoHuJo,HuLoJo,HaJeJoHu,LoLiHuJo,LiVaSa,BurCevPer1,BurCevPer2,HodNak,NakHod,Hod,MetChr,SouHal,BurCevPer3,Min} and references therein. The present work is mainly motivated by the recent 
results of \cite{BurCevPer3,HodNak}.

In this toy model, a population of $N$ agents with limited information and capabilities repeatedly 
compete for a limited global resource, or to be in the minority. 
The desire to be in a minority group is found in many real 
life situations, such as: financial markets, traffic jams, or among 
a group of predators (who wish to hunt in areas with 
fewer competitors).

At each round of the game, every individual has to 
choose whether to be in room `0' (e.g., choosing to sell an asset) 
or in room `1' (e.g., choosing to buy an asset). 
At the end of each turn, agents belonging to the 
smaller group (the minority) are 
the winners, each of them gains $R$ points (the ``prize''), 
whereas the others lose a point (the ``fine''). 
The agents share a common look-up table, containing the outcomes of
recent occurrences. This allows the determination of a ``predicted trend'' in the system, 
which is followed by each agent with probability $p$, known as the agent's ``gene'' value.

In the evolutionary formulation of the model (EMG) agents are allowed 
to evolve (``mutate'') their strategies based on past experience. 
If an agent score falls below some value $d$, he mutates -- 
its gene value is modified. In this sense, 
each agent tries to learn from his past mistakes, 
and to adjust his strategy in order to perform better.

A remarkable conclusion deduced from the EMG \cite{John1} 
is that a population of competing agents tends to {\it self-segregate} into opposing 
groups characterized by extreme behavior. 
It was realized that in order to flourish in such situations, an agent 
should behave in an extreme way ($p=0$ or $p=1$) \cite {John1}. 
On the other hand, in many real life situations the prize-to-fine 
ratio may take a variety of different values \cite{HodNak,BurCevPer1}. A different kind of 
strategy may be more favorite in such situations. 
In recent studies it was found \cite{HodNak} that an intriguing phase transition exist in the model: 
``confusion'' and ``indecisiveness'' take over when the 
prize-to-fine ratio falls below some critical value, 
in which case agents characterized by a coin-tossing strategy ($p={1 \over 2}$) 
perform better than extreme ones. In such circumstances agents tend to 
{\it cluster} around $p={1 \over 2}$ (see Fig. 1 of Ref. \cite{HodNak}) rather 
than self-segregate into two opposing groups.

In \cite{HodNak} we have considered a {\it uniform} strategy updating rule in which 
the new strategy (of a mutating agent) 
is chosen uniformly within the range $0 \leq p \leq 1$. 
Burgus, Ceva and Perazzo \cite{BurCevPer3} have recently considered the 
same model problem, with an updating rule of the 
form $p \to p \pm \delta p$, where  $\delta p < {1 \over 2}$, and found 
that the population tends to form an M-shaped strategy distribution in the $R<1$ case. 
In the present work we further explore this system, and provide some new insights that extend 
and link the results of \cite{HodNak} to those of \cite{BurCevPer3}.

\begin{figure}[tbh]
\centerline{\epsfxsize=9cm \epsfbox{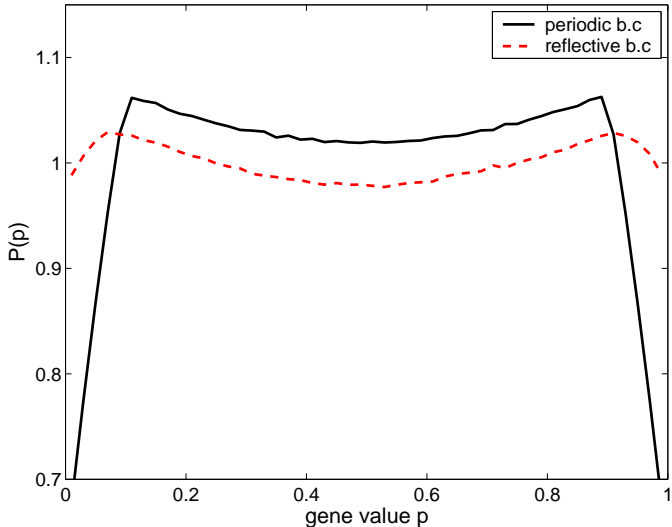}} 
\caption{The strategy distribution $P(p)$ for periodic boundary 
conditions (solid line) and reflective boundary conditions (dashed 
line). The results are for $N=10001$ agents, $R=0.8, d=-4$, 
and $\delta p=0.1$. 
Each point represents 
an average value over 10 runs and 100000 time steps per run.}
\label{Fig1}
\end{figure}

First, we would like to stress the importance of the chosen 
{\it boundary conditions} in the case of an updating rule of the form $p \to p \pm \delta p$ \cite{John1}. 
Figure \ref{Fig1} displays the long-time averaged 
gene distribution $P(p)$ of the agents for two different types of 
boundary conditions: periodic and reflective. One finds that for 
periodic boundary conditions the population tends to cluster 
at intermediate gene values. The curve between the two peaks, 
located at $p=\delta p$ and $p=1-\delta p$, is almost {\it flat}, while 
agents with extreme gene values ($p \simeq 0$ and $p \simeq 1$) 
perform much worse (we shall shortly demonstrate that the gene distribution may also have an 
inverse-U shape, depending on the precise values of $R$ and $\delta p$). 
On the other hand, the gene distribution is almost flat for reflective 
boundary conditions. 

\begin{figure}[tbh]
\centerline{\epsfxsize=9cm \epsfbox{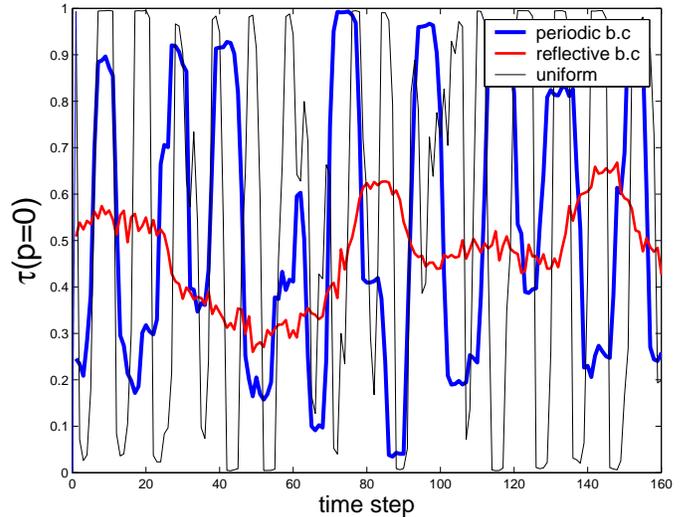}} 
\caption{Temporal dependence of the winning probabilities $\tau(p=0)$, for 
three distinct cases: 
(i)  $\delta p=0.1$ with periodic boundary conditions, 
(ii) $\delta p=0.1$ with reflective boundary conditions, and (iii) uniform updating rule. 
The results are for $N=10001$ agents, $R=0.8$, and $d=-4$.}
\label{Fig2}
\end{figure}

The underlying mechanism which is responsible for this important difference are 
the temporal oscillations observed in the winning probabilities of the agents \cite{HodNak,NakHod}. 
Figure \ref{Fig2} displays the time dependence of the winning probability of a $p=0$ agent 
(the winning probability of a central agent, with $p={1 \over 2}$, is practically constant in time). 
We consider three distinct cases, characterized by: 
(i)  $\delta p=0.1$ with periodic boundary conditions, 
(ii) $\delta p=0.1$ with reflective boundary conditions, and (iii) uniform updating rule. 
One finds smaller oscillation amplitudes and longer periods for reflective boundary 
conditions, as compared to the case of periodic boundary conditions. 
This implies that, for reflective boundary conditions the performance of 
extreme agents ($p=0$ and $p=1$) becomes quite similar to the performance of 
central agents (characterized by $p={1 \over 2}$), implying a flatter gene distribution for these 
boundary conditions. 
On the other hand, for periodic boundary conditions one finds that the temporal oscillations are 
much more similar to the uniform case studied in \cite{HodNak,NakHod} 
(as compared to the case of reflective boundary conditions). 
Indeed, the ratio $P({1 \over 2}):P(0)$ for periodic boundary conditions is very 
similar to the corresponding ratio in the uniform case (compare Fig. \ref{Fig1} with Fig. 1 of 
\cite{HodNak}).

\begin{figure}[tbh]
\centerline{\epsfxsize=9cm \epsfbox{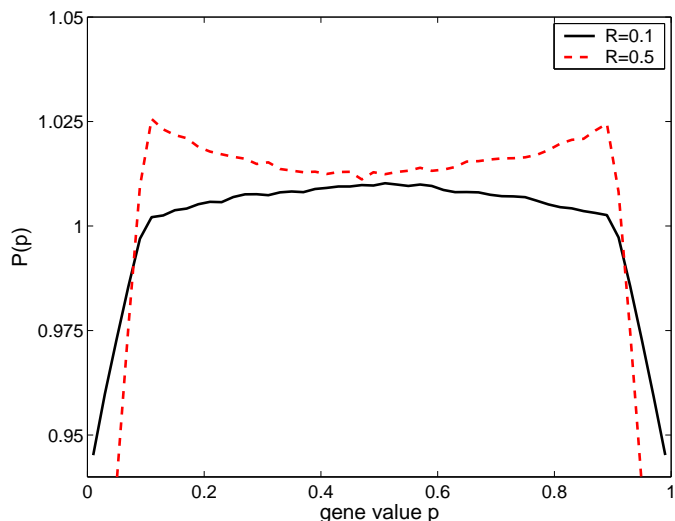}} 
\caption{The strategy distribution $P(p)$ for different values of the 
prize to fine ratio: $R=0.1$ and $R=0.5$. 
The results are for $N=10001$ agents, $d=-4, \delta p=0.1$, and periodic 
boundary conditions. Each point represents 
an average value over 10 runs and 100000 time steps per run.}
\label{Fig3}
\end{figure}

Figure \ref{Fig3} shows the strategy distribution of the population for different prize-to-fine ratios, and 
with $\delta p \ll 1$. 
The results demonstrate the existence of a stable phase characterized by 
an inverse-U shaped gene distribution. However, unlike the uniform case \cite{HodNak}, 
the critical value of $R$ which 
separates the inverse-U distribution from the M-shaped one does not equal $1$ (in the $N \to \infty$ limit).

\begin{figure}[tbh]
\centerline{\epsfxsize=9cm \epsfbox{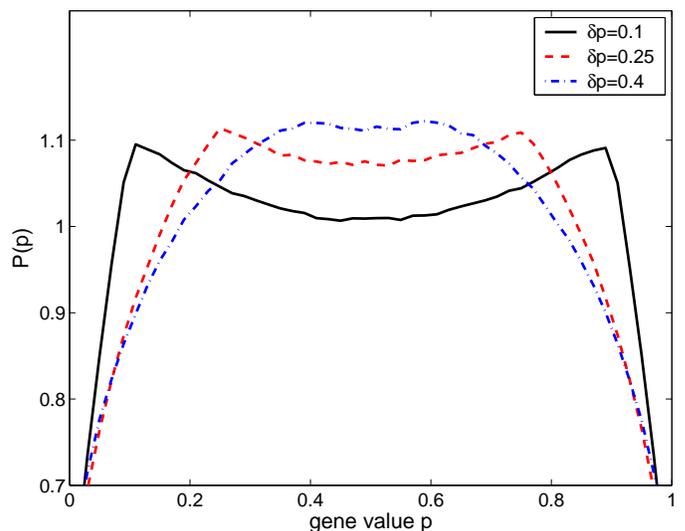}} 
\caption{The strategy distribution $P(p)$ for different $\delta p$ 
values: $\delta p=0.1, 0.25,$ and $0.4$. 
The results are for $N=10001$ agents, $R=0.9, d=-4$, and periodic 
boundary conditions. Each point represents 
an average value over 10 runs and 100000 time steps per run.}
\label{Fig4}
\end{figure}

\begin{figure}[tbh]
\centerline{\epsfxsize=9cm \epsfbox{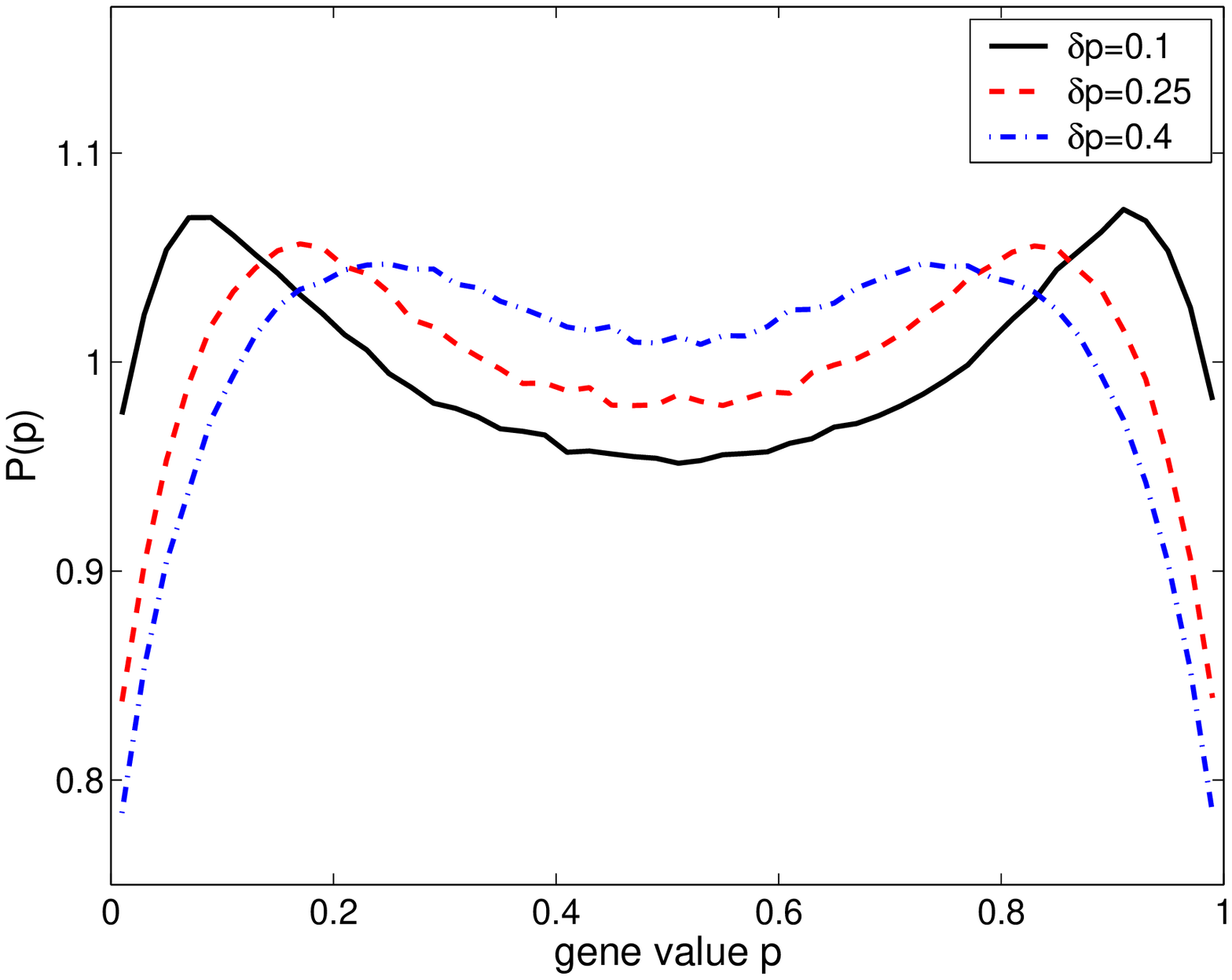}} 
\caption{The strategy distribution $P(p)$ for different $\delta p$ 
values: $\delta p=0.1, 0.25,$ and $0.4$. 
The results are for $N=10001$ agents, $R=0.9, d=-4$, and reflective 
boundary conditions. Each point represents 
an average value over 10 runs and 100000 time steps per run.}
\label{Fig5}
\end{figure}

In Fig. \ref{Fig4} we display $P(p)$ for various different $\delta p$ values with periodic 
boundary conditions. We find that the peaks 
of the strategy distribution (for prize-to-fine ratios which are large 
enough to allow an M-shaped gene distribution) occurs at $p=\delta p$ and its symmetric 
counterpart $1-\delta p$. Regardless of the value of $\delta p$, the agents do {\it not} 
self segregate -- the extreme strategies ($p=0$ and $p=1$) perform worst. 
The strategy distribution moves smoothly into an inverse-U shape 
in the limit of $\delta p={1 \over 2}$ \cite{HodNak}. 
Figure \ref{Fig5} displays the same results for reflective boundary conditions, where 
$\delta p=1$ is equivalent to the uniform updating rule \cite{HodNak}.

\begin{figure}[tbh]
\centerline{\epsfxsize=9cm \epsfbox{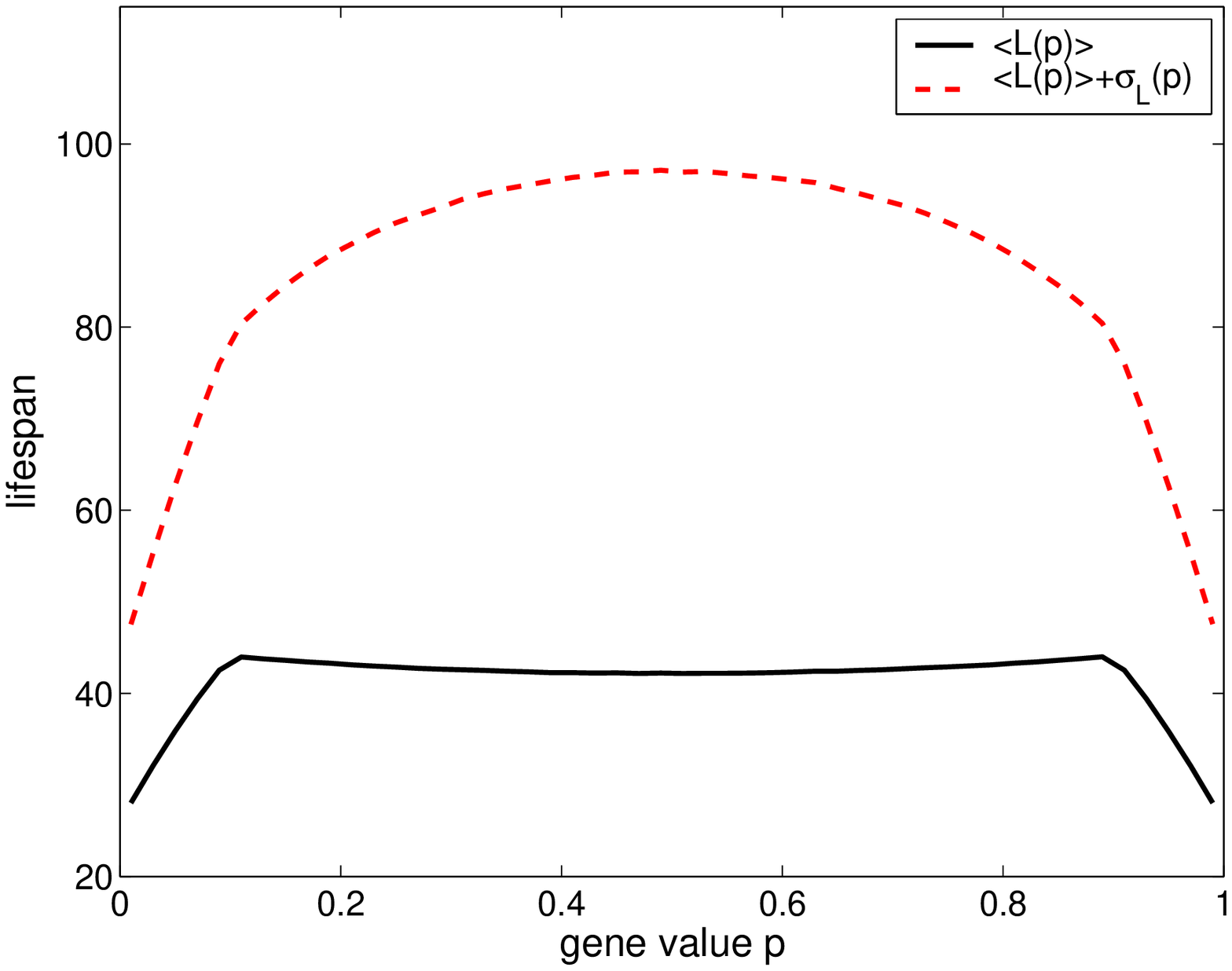}} 
\caption{The average lifespan $<$$L(p)$$>$ (solid curve) and 
$<$$L(p)$$>$+$\sigma_L(p)$ (dashed curve) of the agents. 
The results are for $N=10001$ agents, 
$R=0.8, d=-4, \delta p=0.1$, and periodic boundary conditions. 
Each point represents 
an average value over 10 runs and 100000 time steps per run.}
\label{Fig6}
\end{figure}

Figure \ref{Fig6} displays the average lifespan $<$$L(p)$$>$ of the 
agents. In order to get a better picture 
of the lifespan distribution, we also plot $<$$L(p)$$>$+$\sigma_L(p)$ as a 
function of the gene value $p$. Here $\sigma_L(p)$ is the 
root mean square separation of the lifespans. 
In this case, one finds an 
inverse-U shaped distribution (with the peak occurs at 
$p={1 \over 2}$). This implies that agents characterized by 
$p={1 \over 2}$ (a coin-tossing strategy) have the best chances of 
survival at asymptotically long times, as predicted analytically in 
\cite{Hod}. This important feature is explained by the global 
currents in the gene-space, which {\it reduce} the value of $\sigma_L(p=0)$, and have a 
negligible effect on $\sigma_L(p={1 \over 2})$ \cite{NakHod,Hod}. We emphasize that 
these results hold true for both periodic and reflective boundary conditions.

\begin{figure}[tbh]
\centerline{\epsfxsize=9cm \epsfbox{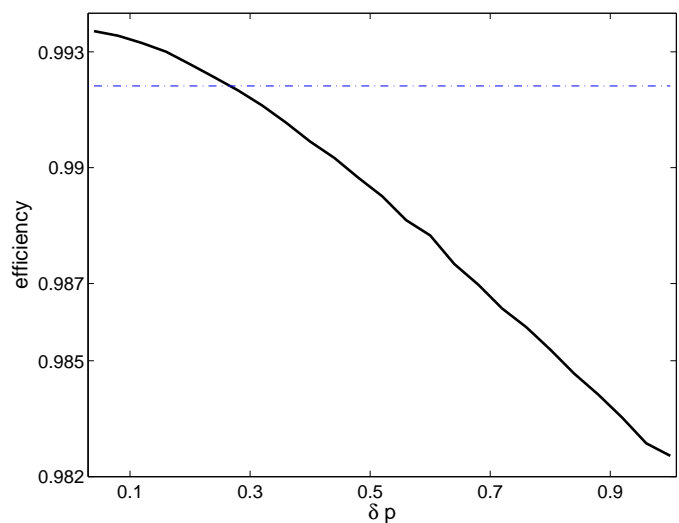}} 
\caption{The efficiency of the system as a function of the length scale, $\delta p$. 
Horizontal line represents the efficiency for a coin-tossing situation. 
The results are for $N=10001$ agents, 
$R=0.7, d=-4$, and reflective boundary conditions.}
\label{Fig7}
\end{figure}

The efficiency of the system is defined as the number of agents in the 
minority room, divided by the maximal possible size of the 
minority group, $(N-1)/2$. 
Figure \ref{Fig7} displays the efficiency as a function of the length scale $\delta p$. 
The system's efficiency is a monotonic decreasing function of $\delta p$. This is caused by the 
fact that larger $\delta p$ values imply {\it larger} temporal oscillations in the occupation numbers of 
the rooms, thus decreasing the number of agents in the winning group 
(and increasing the number of agents in the losing room).

Finally, we would like to address the last point raised in \cite{BurCevPer3}. 
It is claimed that the fluctuations in the average gene 
value $<$$p$$>$ have been considered in \cite{BurCevPer2}. 
However, the {\it oscillatory} behavior of $<$$p$$>$, which is an highly important feature of the system's dynamics was 
{\it not} observed in \cite{BurCevPer2}. Rather, Burgos et. al. \cite{BurCevPer2} find a non-oscillatory value for 
$<$$p$$>$-${1 \over 2}$, see Eq. $(15)$ of \cite{BurCevPer2}. We have shown, on the other hand, that 
the quantity $<$$p$$>$-${1 \over 2}$ displays temporal oscillations, with {\it well defined} frequency and 
amplitude \cite{HodNak,NakHod}. It is important to distinguish between {\it regular} temporal oscillations of the 
physical quantities (such as  $<$$p$$>$) discussed in \cite{HodNak,NakHod}, as opposed to 
thermal fluctuations discussed in \cite{BurCevPer2}. 
Thermal fluctuations of a thermodynamic system are essentially random in nature, whereas we have found 
regular oscillations, which are characterized by well defined frequency and amplitude. The 
{\it oscillatory} nature of $<$$p$$>$ \cite{HodNak,NakHod} 
has been proven to be an essential feature which is responsible for 
the dynamical phase transition (from self-segregation to clustering) observed in the EMG \cite{Hod}.
We would like to emphasize that these oscillations exist also for complex systems with 
a strategy updating rule of the form 
$p \to  p \pm \delta p$, regardless of the value of $\delta p$ and 
the type of boundary conditions used (see. Fig. \ref{Fig2}). 

\bigskip
\noindent
{\bf ACKNOWLEDGMENTS}
\bigskip

SH thanks a support by the 
Dr. Robert G. Picard fund in physics. 
This research was supported by grant 159/99-3 from the Israel Science Foundation. 
EN thanks a support by the Horwitz foundation.


\begin{thebibliography}{99}

\bibitem{HubLuk} B. Huberman and R. Lukose, Science {\bf 277}, 535 (1997).

\bibitem{ChaZha} D. Challet and C. Zhang, Physica A {\bf 246}, 407 (1997); 
{\bf 256}, 514 (1998); {\bf 269}, 30 (1999).

\bibitem{DhRo} R. D`Hulst and G. J. Rodgers, Physica A {\bf 270}, 514 (1999).

\bibitem{SaMaRi} R. Savit, R. Manuca and R. Riolo, Phys. Rev. Lett. {\bf 82}, 2203 (1999).

\bibitem{John1} N. F. Johnson, P. M. Hui, R. Jonson, and T. S. Lo, Phys. Rev. 
Lett. {\bf 82}, 3360 (1999).

\bibitem{Cev} H. Ceva, Physica A {\bf 277}, 496 (2000).

\bibitem{BurCev} E. Burgos and H. Ceva, Physica A {\bf 284}, 489 (2000). 

\bibitem{LoHuJo} T. S. Lo, P. M. Hui and N. F. Johnson, Phys. Rev. E {\bf 62}, 4393 (2000).

\bibitem{HuLoJo}  P. M. Hui, T. S. Lo, and N. F. Johnson, e-print cond-mat/0003309.

\bibitem{HaJeJoHu} M. Hart, P. Jefferies, N. F. Johnson and P. M. Hui, e-print cond-mat/0003486; Phys. 
Rev. E {\bf 63}, 017102 (2000).

\bibitem{LoLiHuJo} T. S. Lo, S. W. Lim, P. M. Hui and N. F. Johnson, Physica A {\bf 287}, 313 (2000).

\bibitem{LiVaSa} Y. Li, A. VanDeemen and R. Savit, e-print nlin.AO/0002004.

\bibitem{BurCevPer1} E. Burgos, H. Ceva and R. P. J. Perazzo, Physica A {\bf 294}, 539 (2001); 
Phys. Rev. E {\bf 64}, 016130 (2001); e-print cond-mat/0212635.

\bibitem{BurCevPer2} E. Burgos, H. Ceva and R. P. J. Perazzo, Phys. Rev. E {\bf 65}, 036711 (2002).

\bibitem{HodNak} S. Hod and E. Nakar, Phys. Rev. Lett. {\bf 88}, 238702 (2002).

\bibitem{NakHod} E. Nakar and S. Hod, Phys. Rev. E {\bf 67}, 016109 (2003).

\bibitem{Hod} S. Hod, e-print cond-mat/0212055.

\bibitem{MetChr} R. Metzler and  C. Horn, e-print cond-mat/0212481.

\bibitem{SouHal} A. Soulier and T. Halpin Healy, e-print cond-mat/0209451.

\bibitem{BurCevPer3} E. Burgos, H. Ceva and R. P. J. Perazzo, e-print cond-mat/0301518.

\bibitem{Min} http://www.unifr.ch/econophysics/minority.

\end{thebibliography}
\end{document}